\documentclass[iop]{emulateapj}
\usepackage{amsmath,amssymb,graphicx,longtable,booktabs}

\newcommand{\sersic}{Sersi\'c}
\newcommand{\magarcsec}{mag arcsec$^{-2}$}	

\newcommand{\gclfmu}{27.53}
\newcommand{\gclfmuerr}{0.34}
\newcommand{\gclfsig}{0.76}
\newcommand{\gclfsigerr}{0.23}
\newcommand{\gcnum}{28}
\newcommand{\gcnumerr}{14}
\newcommand{\gcsn}{26}
\newcommand{\gcsnerr}{13}
\newcommand{\sersicn}{0.85}
\newcommand{\sersicnerr}{0.22}
\newcommand{\sersicre}{12}
\newcommand{\sersicreerr}{2}

\slugcomment{Accepted for publication in The Astrophysical Journal Letters}

\shorttitle{A Rich Globular Cluster System in an Ultra-Diffuse Galaxy}
\shortauthors{Peng \& Lim}

\begin{document}


\title{A Rich Globular Cluster System in Dragonfly 17: \\ 
	Are Ultra-Diffuse Galaxies Pure Stellar Halos?} 


\author{Eric W. Peng\altaffilmark{1,2}}

\author{Sungsoon Lim\altaffilmark{1,2}}

\altaffiltext{1}{Department of Astronomy, Peking University, Beijing
  100871, China; peng@pku.edu.cn}
\altaffiltext{2}{Kavli Institute for Astronomy and Astrophysics, Peking
  University, Beijing 100871, China}


\begin{abstract}

Observations of nearby galaxy clusters at low surface
brightness have identified galaxies with low luminosities, but sizes
as large as $L^\star$ galaxies, leading them to be
dubbed ``ultra-diffuse galaxies'' (UDGs). The survival of
UDGs in dense environments like the Coma cluster suggests that UDGs
could reside in much more massive dark halos. We report the detection of a substantial population of
globular clusters (GCs) around a Coma UDG, Dragonfly 17 (DF17). We find that DF17 has a high GC
specific frequency of $S_N=\gcsn\pm\gcsnerr$. The GC system is
extended, with an effective radius of $\sersicre\arcsec\pm2\arcsec$, or $5.6\pm0.9$~kpc at
Coma distance, $70\%$ larger than the galaxy itself.
We also estimate the mean of the GC luminosity function to infer a
distance of $97^{+17}_{-14}$~Mpc, providing redshift-independent confirmation that one of these UDGs is in the Coma
cluster. The presence of a rich GC system in DF17 indicates that,
despite its low stellar density, star formation was intense enough to form many massive star
clusters. If DF17's ratio of total GC mass to total halo mass is similar to those in other galaxies, then DF17 has an inferred total
mass of $\sim\!10^{11} M_\odot$, only $\sim\!10\%$ the mass of the Milky Way,
but extremely dominated by dark matter, with $M/L_V\approx1000$. We suggest that UDGs like DF17 may be ``pure stellar halos'', i.e., galaxies that formed their stellar halo components, but then suffered an early cessation in star formation that prevented the formation of any substantial central disk or bulge.

\end{abstract}




\section{Introduction}
\setcounter{footnote}{0}

The realm of low surface brightness is still one of the most
unexplored in astronomy. Many of the visible counterparts to
structures expected in a cold dark matter Universe are expected to be
at a surface brightness much fainter than that of the night sky. A
recent study using a new telescope optimized for low surface 
brightness imaging, the Dragonfly Telephoto Array, in conjunction with
imaging from the Canada-France-Hawaii Telescope (CFHT), reported the
discovery of 47 UDGs ($\mu_{(e,g)}=24$--$26$~\magarcsec) in the
direction of the Coma cluster of galaxies \citep{vanDokkum15a}. These galaxies have
luminosities and appearances similar to early-type dwarf galaxies, but
have much larger sizes, similar to $L^\star$ galaxies
($1.5<R_e<4.5$~kpc). Their luminosities and sizes make them outliers
in traditional scaling relation diagrams for galaxies. One UDG, DF44, has been
confirmed to have a redshift consistent with membership in the Coma
cluster \citep{vanDokkum15b}. Subsequently, 854 new diffuse galaxies have been
identified in Coma \citep{Koda15}, and
even lower surface brightness galaxies have been identified in the
nearby Virgo cluster of galaxies \citep{Mihos15,Beasley16} showing
that these galaxies are far from uncommon, and that they may even be able to
exist in the dense core of a galaxy cluster. Unfortunately, their low
surface brightness and apparent lack of ongoing star formation makes
it prohibitively difficult to study their stellar
populations. Moreover, the distance of the Coma cluster \citep[$\approx100$~Mpc;][]{Carter08} puts resolved studies of their red giant branch (RGB) stars
beyond the capabilities of current telescopes.   

One probe of stellar populations that is within reach for these
diffuse galaxies is globular clusters (GCs). GCs are old, compact star
clusters whose presence in galaxies points to an early
epoch of galaxy building where the intense star formation needed to
form massive star clusters was commonplace. Observationally, they are
useful tracers of old stellar populations because they can be observed
at large distances. The number of GCs in a galaxy also
correlates linearly with the total host halo mass \citep{Blakeslee97,Peng08,Spitler09,Harris13}, giving a way to
estimate the total mass of a galaxy without other information. 
A recent study by \citet{Beasley16} used kinematics of GCs in the Virgo 
cluster UDG VCC 1287 to show that the galaxy has an extremely high mass-to-light
ratio, consistent with the premise that the number of GCs trace total
mass, even in these extreme galaxies. The GC
luminosity function peaks at a nearly universal luminosity, allowing an
estimate of the distance to the galaxy. At the distance of the Coma
cluster, GCs are readily visible in Hubble Space Telescope ({\it HST})
images. 

We use archival {\it HST} Advanced Camera for Surveys Wide Field
Channel (ACS/WFC) imaging of one Coma UDG, Dragonfly 17 (DF17), to investigate
whether UDGs in the dense Coma cluster can host a system of GCs, and what that implies for the mass and origin of UDGs.

\begin{figure*}
\plotone{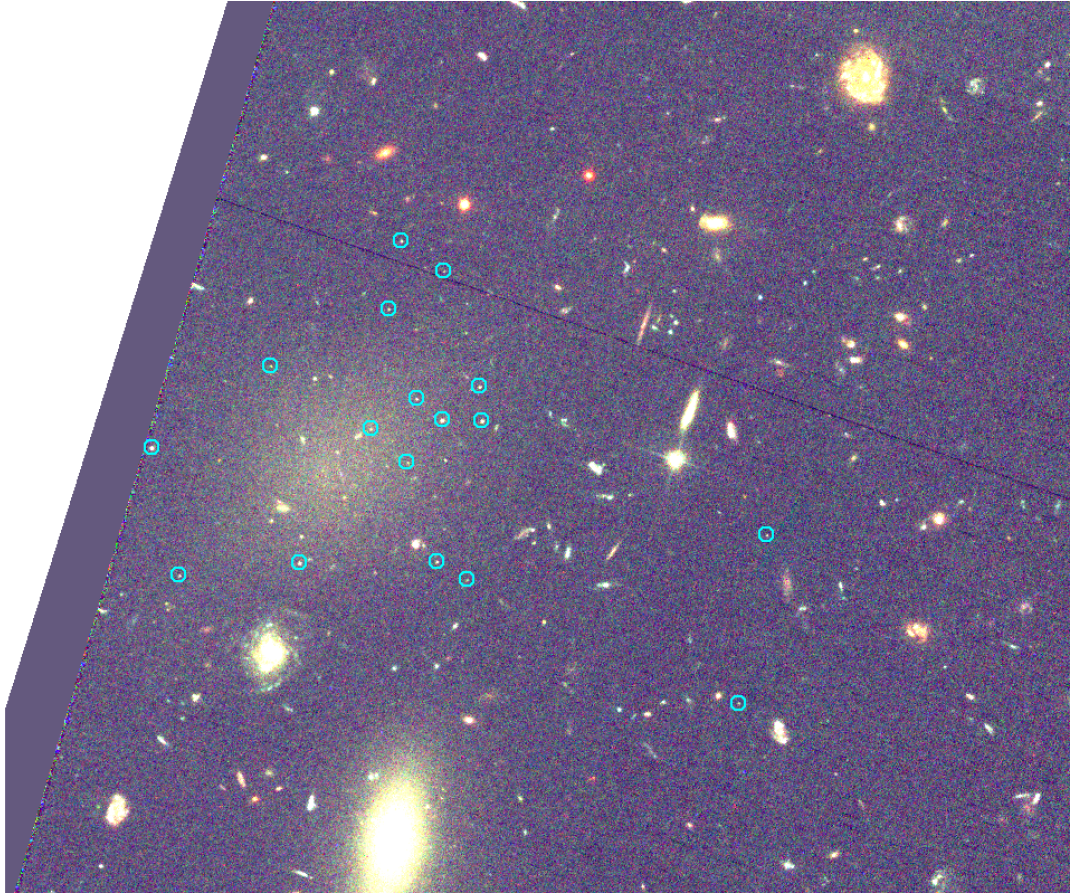}
\caption{This $68\arcsec\times 57\arcsec$ image shows DF17 on the edge of the
HST/ACS field of view. GC
candidates (cyan circles) selected by morphology and color are clustered around the
galaxy. The GCs roughly follow the star
light, although there is some lopsidedness in the GCs,
with nearly three times as many GCs on the NW side along the major
axis than on the SE. There is no obvious evidence for a nuclear star
cluster. North is up, and East is left. \label{fig:image}} 
\end{figure*}
 
\pagebreak
\section{Observations and Globular Cluster Candidate Selection}

We used images taken with the {\it HST} {\it ACS/WFC} (GO-12476, PI: Cook,
\citealt{Macri13}). These data were described in
\citep{vanDokkum15a}, and are comprised of parallel
observations in three filters ($g_{475}$, $V_{606}$, and
$I_{814}$). We made deep stacked images using the AstroDrizzle
task in DrizzlePac\footnote{http://www.stsci.edu/hst/HST\_overview/drizzlepac}. Total
accumulated exposure times are 5100s, 5820s, and 5100s for the F475W,
F606W, and F814W images, respectively. The pixel scale of the
output images is $0\farcs05/$pix, and the full width at half maximum
(FWHM) of point sources in the images is about two pixels. 

We generated an object catalog using the Source Extractor
\citep{Bertin96} software package. For the detection image, we used a signal-to-noise ($S/N$)-weighted, combined image of all three filters. We performed aperture photometry within a $0\farcs3$
diameter aperture, derived the aperture correction to a $0\farcs5$
aperture using bright stars in the field, and then applied the
aperture corrections to $5\farcs5$ from \citet{Bohlin11}. 

GCs in the Coma cluster are 
unresolved, even with $HST$, so we first identified GC candidates as point
sources. Similar to \citet{Peng11}, we defined an ``inverse concentration''
index ($C_{4-7}$), which is the difference in magnitude between a 4-pixel 
aperture ($0\farcs2$) and a 7-pixel aperture ($0\farcs35$),
normalized so that for point sources, $C_{4-7}=0$.  Although we detect objects as faint as $V_{606}\approx30$~mag, objects
with $V_{606}\gtrsim29$~mag are overwhelmingly background sources, as
the number counts of distant galaxies increases exponentially. For
this reason, we define two GC candidate samples. The first is
an ``optimal'' sample, which minimizes contamination and
increases contrast between any GC population and the background. This
sample has a flux limit of $V_{606}<28.25$~mag, and becomes
increasingly stringent on the concentration criterion (removing
more extended sources) for the faintest 2~mag. 
Our second sample is the ``deep'' sample, which has a
simple flux limit of $V_{606}<29$~mag with symmetric cuts at
$\pm0.1$~mag in concentration. 

For both samples, we used the
($g_{475}-V_{606}$) and ($V_{606}-I_{814}$) colors to
identify GC candidates by comparing objects in our field to the colors
of GCs in the Virgo central galaxy, M87, which have HST/ACS photometry
in the exact same filters \citep{Peng09, Jordan09}. We used
the M87 GCs to define a locus in color space, and selected as
GCs all objects whose $2\sigma$ photometric uncertainty ellipse
included the GC locus. This allows for
more stringent selection for objects with more precise colors. 
We use the ``optimal'' sample for all subsequent
analyses except for the measurement of the GC luminosity function
(Section~\ref{sec:gclf}). 

\section{Results}

\subsection{The GC system of DF17}
Figure~\ref{fig:image} shows an image of DF17 with GC candidates
from the ``optimal'' sample circled. The
spatial distribution of GC candidates is highly concentrated toward
DF17. The GC candidates roughly follow the stellar light of the galaxy, with a steep 
drop in numbers just beyond DF17's optical radius. This
strongly suggests that the GC candidates are physically associated with
DF17. Within
$20\arcsec$ of the center of DF17 (a radius which contains all of the GC
candidates clustered around the galaxy), we expect only 1.4 of the 15 selected objects to be background contaminants. This indicates that nearly all of the GC candidates around DF17 are, in fact, real GCs associated with the galaxy.

The GCs appear to be asymmetrically distributed about the galaxy, with
nearly three times as many on the NW side along the major axis (which
runs roughly NW-SE). Because DF17 is on the edge of the field of view,
it is difficult to say whether this imbalance is mitigated at larger
radii to the SE. We also note that there is no obvious nuclear star
cluster above our detection limit. 

\begin{figure}
\epsscale{1.2}
\plotone{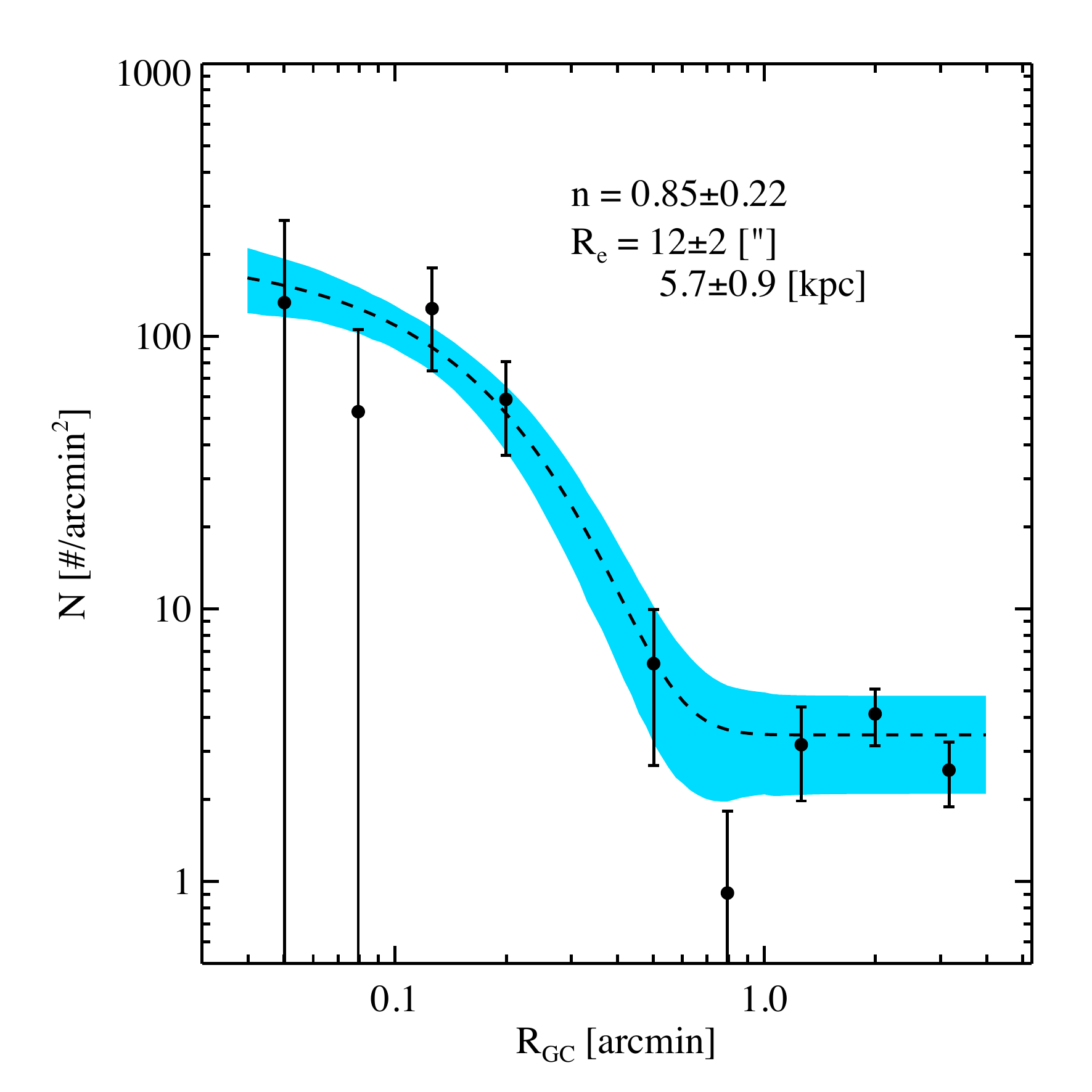}
\caption{The binned radial surface density profile of GCs in DF17 with best-fit
  \sersic\ profile overplotted. The profile has a flat
  core followed by a steep drop in the outer regions, which is
  consistent with a low \sersic\ index. The maximum likelihood
  estimates for the \sersic\ function parameters are $n=\sersicn\pm\sersicnerr$ and
  $R_e=\sersicre\arcsec\pm\sersicreerr\arcsec$. The fit was performed
  directly on the unbinned data. For comparison, the
  \sersic\ fit for the stellar light in DF17 is $n=0.6$ and
  $R_{e,stars}=7\farcs0$.\label{fig:radial}} 
\end{figure}

Figure~\ref{fig:radial} shows the GC surface density profile. We characterize the profile using 
a  \sersic\ function and constant background, with best-fit parameters estimated using maximum likelihood and bootstrap resampling. We find that the GC system has
a profile consistent with a low \sersic\ index, $n=\sersicn\pm\sersicnerr$, with
a flat core and a steep outer drop off. The size of the GC system is
characterized by the effective radius, $R_{e,GC}=\sersicre\arcsec\pm\sersicreerr\arcsec$. For comparison, the
effective radius measured for the stars from the same imaging is $R_{e,stars}=7\farcs0$ \citep{vanDokkum15a}, so the GCs have an extent $70\%$ larger than the already diffusely distributed stars, something which is also seen in more massive galaxies \citep[e.g.,][]{McLaughlin99,Kartha14}.

\subsection{The GC Luminosity Function and the Distance to DF17}
\label{sec:gclf}

\begin{figure}
\epsscale{1.2}
\plotone{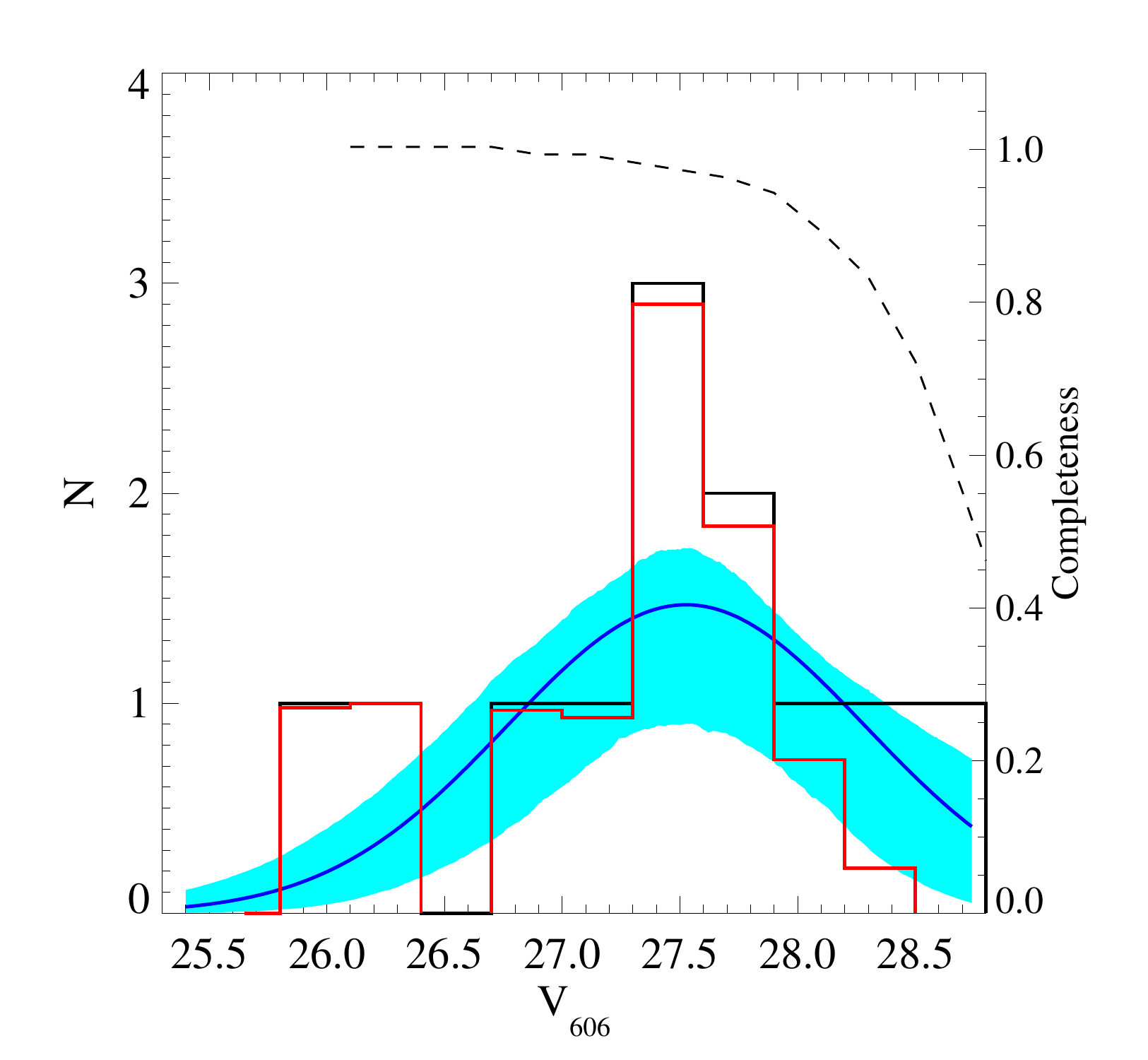}
\caption{The GC luminosity function in DF17. We show the binned
  number counts of GC candidates (black histogram) and
  background-subtracted GC counts (red histogram) from our ``deep''  
  sample within the central $12\arcsec$ ($1 R_{e,GC}$) of DF17. Overplotted is the
  best-fit Gaussian GCLF with $\mu_{V606}=\gclfmu\pm\gclfmuerr$ 
  and $\sigma=\gclfsig\pm\gclfsigerr$ (blue line) and the
  68\% confidence regions determined from the bootstrap
  (cyan). We also show the completeness of our
  observations as a function of $V_{606}$ (dashed line). The
  observations are 90\% complete $1\sigma$ fainter than the GCLF
  peak. Assuming the peak of the GCLF in DF17 is 
  similar to that in other nearby galaxies, DF17 is at a distance of 
  $97^{+17}_{-14}$~Mpc. The fit was performed on the unbinned data.\label{fig:gclf}}
\end{figure}

The luminosity of the GCLF peak is a well-known distance indicator \cite[e.g.,][]{Harris01}, and
measuring its apparent magnitude provides perhaps the only way to obtain a redshift-independent distance for such a distant UDG. At the
distance of the Coma cluster ($m-M\approx35$), a GC
system with a standard Gaussian luminosity function for a low-mass early-type galaxy (with mean $M_V=-7.3$~mag;
\citealt{Miller07}) has a peak whose measurement is within the reach of these 
observations ($V_0\approx27.7$~mag). \citet{Lee16} showed that GCLF distances to the Coma cluster can be obtained with HST imaging. For this analysis, we use our ``deep'' sample, which
extends to $V_{606}=29$~mag. 

We used artificial star tests to quantify the detection efficiency
across our images. We used DAOPHOT~II \citep{Stetson87} to construct an
empirical PSF using bright stars.  We added artificial stars to images in
all three filters, giving them a typical GC color, then ran the same
detection and selection procedures as were used
to generate the ``deep'' sample. The 90\% completeness level is $V_{606}=28.1$~mag and the
50\% completeness level is $V_{606}=28.8$~mag. This latter limit is fainter
by $\sim1$~mag than the expected GCLF peak at the distance of
Coma (see the dashed line in Figure~\ref{fig:gclf}).

We used a Gaussian form for the GC luminosity
function, a power law for the number counts of background sources
(distant, unresolved galaxies), and multiplied both by the derived
detection efficiency function. Using maximum likelihood estimation, we fit
the normalization and power law slope of the background counts using
the GC candidates outside a radius of $36\arcsec$ ($3R_{e,GC}$) from DF17, assuming
that all of these objects are background objects. We fixed the
power law model for the background to then estimate the Gaussian GCLF 
parameters for GC candidates within $12\arcsec$, the
effective radius of the GC system. Figure~\ref{fig:gclf} shows the
number counts of GC candidates within the central $12\arcsec$ of DF17,
as well as the background-subtracted GC counts. $HST$ resolution
eliminates much of the background contamination at these magnitudes,
and the small area in which the GCs reside also helps suppress contamination.  

We find the best-fit parameters for the Gaussian GCLF to be
$\mu_{V_{606}}=\gclfmu\pm\gclfmuerr$~mag and 
$\sigma_{V_{606}}=\gclfsig\pm\gclfsigerr$~mag. Figure~\ref{fig:gclf}
shows the best-fit Gaussian model, as well as the region that
encompass 68\% of the models fit using 1000 bootstrap resamples.

Because the $V_{606}$
magnitude is not a standard $V$, we use transformations from
\citet[][Equation 3]{DeGraaff07} to convert the best-fit $\mu_{V_{606}}$ to
$\mu_V$. Using the color of a typical GC, we find that
$V-V_{606}=0.13$~mag, and $A_V=0.025$ \citep{Schlafly11}, which
results in the GCLF mean being
$\mu_{V_0}=27.63\pm0.35$. Assuming that the form of 
DF17's GCLF is similar to those of 
nearby low-mass galaxies, this value of $\mu_{V_0}$ places DF17 at a distance
of $97^{+17}_{-14}$~Mpc, putting it exactly in the Coma cluster.

\subsection{Total Number of GCs and Specific Frequency}

With our assumption of the form of the GCLF and the GC spatial density profile
in DF17, we can estimate the total number of GCs. Counting seven GCs within $R_e$ and brighter than the estimated GCLF mean, we find
that $N_{GC}=\gcnum\pm\gcnumerr$ (where we include the uncertainty in the GCLF mean). DF17 has a luminosity of $M_V=-15.1$~mag 
(using our distance and transforming from $V_{606}$ reported in
\citealt{vanDokkum15a} ), which gives a GC specific frequency of
$S_N=\gcsn\pm\gcsnerr$.

\section{Discussion}
\subsection{The Mass of DF17}

\begin{figure}
\epsscale{1.2}
\plotone{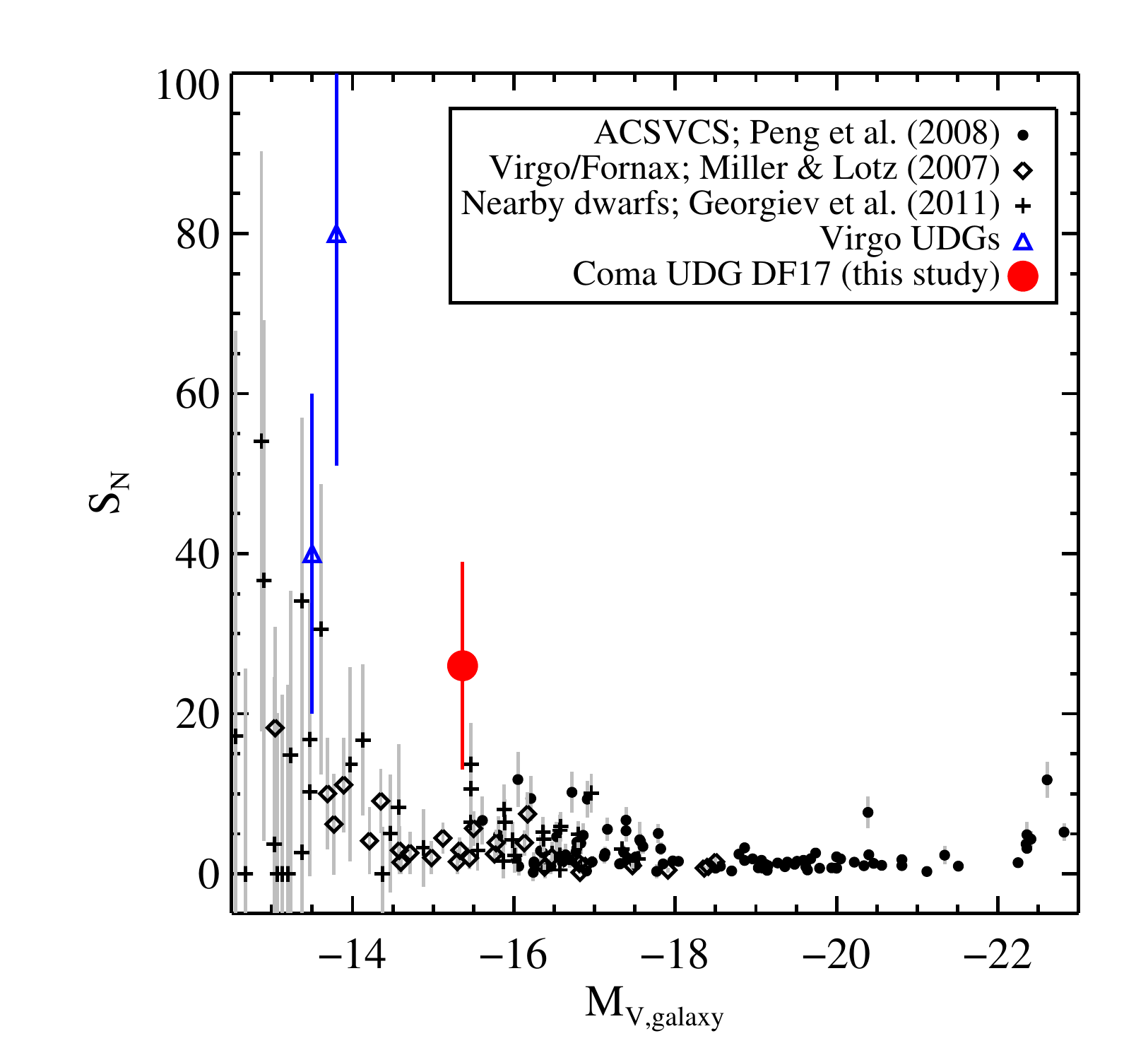}
\caption{GC specific frequencies ($S_N$) as a function of galaxy $M_V$. We plot $S_N$ from three samples of galaxies with relatively high-precision $S_N$ measurements, all derived from $HST$ imaging: the Virgo cluster early-type galaxy sample from the ACSVCS \citep[][dots]{Peng08}, the Virgo and Fornax cluster early-type dwarf galaxy sample of \citet[][crosses]{Miller07}, and the nearby low-mass galaxy sample of \citet[][diamonds]{Georgiev10}. We also plot the high-$S_N$ Virgo cluster ultra-diffuse galaxies from \citet{Mihos15} and \citet{Beasley16} (blue traingles). The Coma UDG from this study, DF17, is the red filled circle. For its luminosity, DF17 has a very high $S_N$, one which is only seen in galaxies at least two magnitudes fainter. \label{fig:sn}}
\end{figure}

DF17 has a substantial population of
GCs, with a specific frequency
($S_N=\gcsn\pm\gcsnerr$) that is among the highest measured for any galaxy,
and is the highest for its 
luminosity range. Figure~\ref{fig:sn} shows a compilation of high-quality, $HST$-derived
$S_N$ values for early-type galaxies (ETGs) across a range of
luminosities. DF17 is a clear outlier at its luminosity, having a value of $S_N$ only seen in galaxies two magnitudes fainter.  Two extreme cases are VCC~1287 \citep{Beasley16}, which has $S_N\sim80$, and another ultra-diffuse galaxy in Virgo \citep{Mihos15} that has $S_N\sim40$, but there are also low-mass galaxies with high $S_N$ from \citet{Georgiev10}. 

Work over the past couple decades has shown that the number of GCs in ETGs
appears to be a reasonable estimator of the total mass of a
galaxy or cluster \citep{Blakeslee97,McLaughlin99}. As a result, the relationship between GC specific frequency
(or GC stellar mass fraction) and galaxy stellar mass traces reasonably well the total
mass--stellar mass relation for galaxies, both with a minimum around
$L^\star$ \citep{Peng08,Spitler09,Harris13,Hudson14}. Provided that this
relation holds across all galaxies, we can use the number of GCs in
DF17 to estimate its total mass. We use the \citet{Harris13} relation between
the GC system mass and the total galaxy mass ($M_{GCS}/M_{halo} =
6\times10^{-5}$) to determine a total mass for DF17 of
$(9.3\pm4.7)\times10^{10} M_\odot$, and a mass-to-light ratio of $M/L_V\approx 1000$. This total mass is roughly $\sim\!10\%$ that of the Milky Way.

The survival of UDGs in dense environments like Coma and other massive clusters \citep{vanderBurg16} provides independent evidence that UDGs must be dark-matter dominated. Moreover, \citet{Beasley16} used GC kinematics to find that the Virgo UDG, VCC~1287, has $M/L\approx 3000$, which was in line with what they inferred from their total number of GCs. These high mass-to-light ratios, and optical colors consistent with old stellar populations, suggest that tremendous gas loss (or lack of accretion) could have occurred early in the evolution of UDGs.

\begin{figure}
\epsscale{1.2}
\plotone{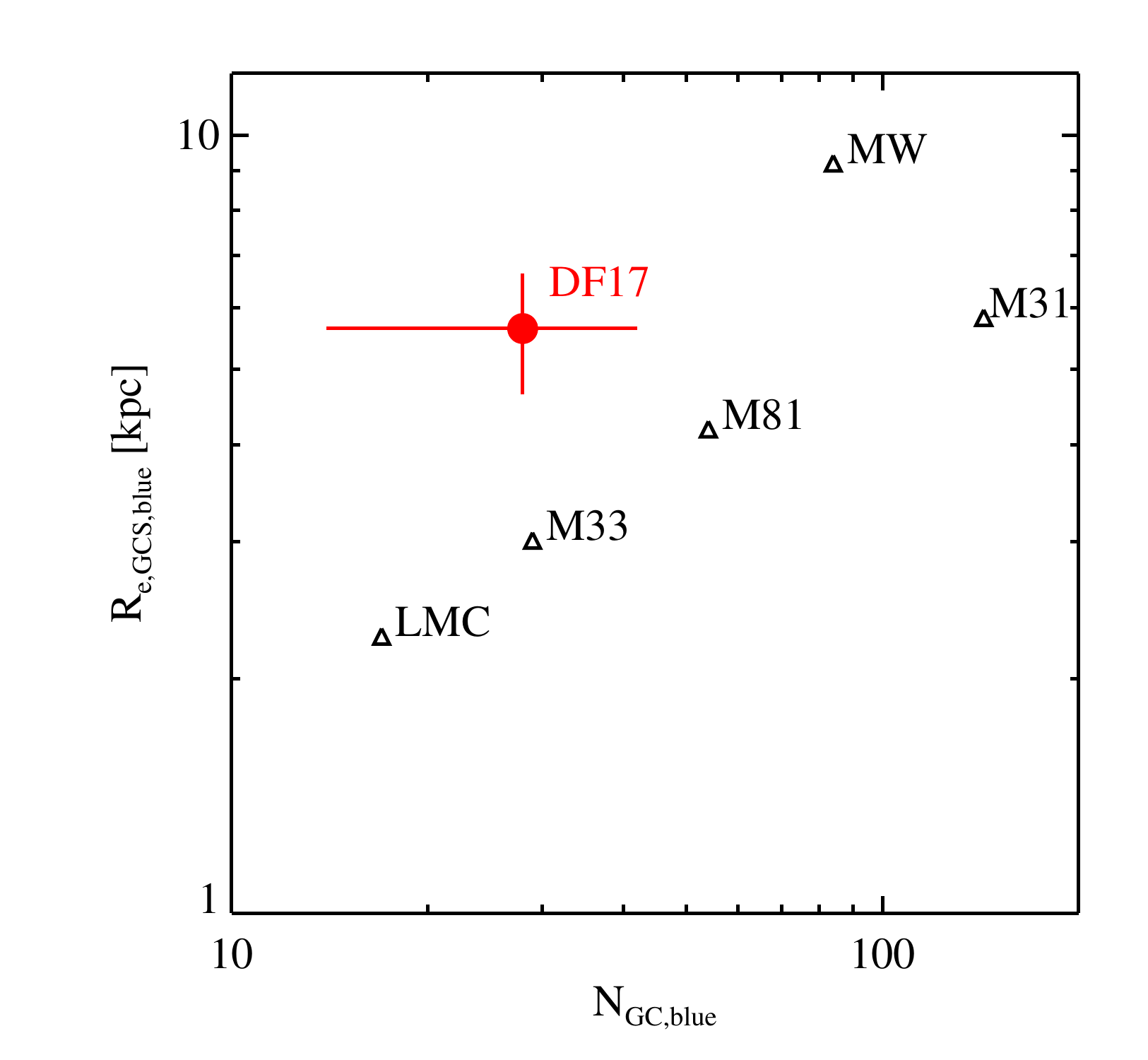}
\caption{The half-number radii ($R_e$) of the GC system versus of the total number of metal-poor (blue) GCs for DF17 compared to Local Group galaxies (Milky Way, M31, M33, and LMC) and the nearby spiral galaxy, M81 \citep{Harris96,Caldwell11,vandenBergh04,SanRoman10,Fan14,Nantais10}. All the GCs in DF17 are assumed to be metal-poor. More massive GC systems are generally larger, and although the uncertainties are likely large for these galaxies, it appears that DF17's GCs, like its stars, are in a more extended distribution than would be expected for its mass. Thus, it's GC system is also ultra-diffuse. \label{fig:reff}}
\end{figure}

\subsection{Ultra-Diffuse Galaxies as ``Pure Stellar Halos''}

The diffuseness of DF17's stellar population seems to be at
odds with the presence of GCs. Star clusters massive enough to be GC progenitors
require intense star formation episodes, like those
seen in gas-rich galaxy mergers. The existence of GCs require this kind of starburst in DF17, but a high star formation rate density should also produce a high surface brightness stellar component. \citet{Amorisco16} suggested that UDGs could be the high angular momentum tail of the normal dwarf galaxy population. While their model produces the proper number of UDGs in clusters, it is unclear how such extended, low surface brightness (LSB) galaxies can produce massive star clusters. Present-day LSB disks, also thought to result from high angular momentum \citep{Dalcanton97}, typically have little molecular gas and low star formation efficiencies \citep{Bothun97}.

One way to explain a high $S_N$ in a UDG is to posit that massive
star clusters preferentially form earlier in a given star formation
episode \citep[e.g.,][]{Peng08}, and then the subsequent formation of field stars is rapidly
quenched. This behavior is supported by measurements of
[$\alpha/Fe$] in Virgo dwarf galaxies \citep{Liu16}, in which dwarfs with higher $S_N$ tend to have higher [$\alpha/Fe$],
suggesting a more rapid (and truncated) star formation history. In this
scenario, DF17 should also have chemical abundance
patterns typical of rapid star formation. 

The constituents of DF17---an extended, spheroidal stellar population and a rich system of GCs---make it most similar to the stellar halos of normal galaxies. In effect, UDGs like DF17 are ``pure stellar halos'': galaxies in which the old stellar halo forms, but then rapid gas removal and ``starvation'' prevents the formation of a more centrally concentrated disk or bulge component. Model decomposition of ETGs reveals the existence of extended, blue components with characteristic sizes of $\sim\!10$~kpc that may be the old stellar halo \citep{Huang13}. Less massive disk galaxies like M33 also have old stellar halos \citep{McConnachie10}. If DF17 were to have a more ``normal'' specific frequency, keeping fixed the number of GCs, it would have a luminosity of $M_V\approx-18$ to $-19$~mag, comparable to that of M33. 

In Figure~\ref{fig:reff}, we compare the effective radius ($R_e$) of the DF17 GC system to those of the metal-poor GC systems for the Milky Way, M31, M33, the LMC, and M81. We derive $R_e$ for these galaxies using catalogs in the published literature \citep{Harris96,Caldwell11,vandenBergh04,SanRoman10,Fan14,Nantais10}. These measurements are uncertain due to the inhomogeneous coverage and depths of the various surveys, but a trend is apparent where metal-poor GC systems with more GCs have larger sizes. DF17's size stands out in this comparison, as its GC system has a size consistent with the more massive galaxies. However, its total number of GCs, and its total stellar mass, is significantly lower than those for the metal-poor stellar halos in the larger galaxies. We note that any subsequent central star formation would likely to move DF17 to the lower right (smaller size, more GCs) in this plot, in the direction of the other galaxies.

While the inferred mass of DF17 makes it unlikely that it is a ``failed'' Milky Way-like galaxy, DF17 could be the stellar halo of something less massive. A bright, relatively compact central stellar component in DF17 might look similar to a ``normal'', sub-$L^\star$ galaxy with a rather extended GC system. We speculate that ultra-diffuse galaxies like DF17 are analogous to the stellar halos of more luminous, high surface brightness galaxies, and that UDGs are part of the surviving population of proto-galaxies that built up the high-$S_N$, high-$M/L$ stellar populations in the outskirts of today's massive galaxies.

Our ability to detect GCs in UDGs out to the distance of the Coma cluster with {\it HST} gives us a relatively inexpensive way to estimate their total masses, distances, and study their star formation histories. Combined with the measurements of the extent of these GC systems, which can be compared to those of nearby galaxies, GCs will be a valuable resource to investigate the origins of UDGs.

\acknowledgments

EWP and SL acknowledge support from the National Natural
Science Foundation of China under Grant No.\ 11573002,
and from the Strategic Priority Research Program, ``The Emergence of
Cosmological Structures'', of the Chinese Academy of Sciences, Grant
No. XDB09000105. Based on observations with the NASA/ESA Hubble 
Space Telescope obtained from the Mikulski Archive for Space Telescopes (MAST) at the Space Telescope Science Institute, which is operated by the Association of Universities for Research in
Astronomy, Inc., under NASA contract NAS 5-26555. This research has made use of the NASA/IPAC Extragalactic Database (NED). This research made use of Astropy, a community-developed core Python package for Astronomy \citep{Astropy13}. 

Facilities: \facility{HST(ACS)}

\bibliographystyle{apj}

 \newcommand{\noop}[1]{}

\clearpage

\end{document}